\documentclass[a4paper]{article}
\usepackage[utf8]{inputenc}
\usepackage[english]{babel}
\usepackage{fullpage}
\usepackage{caption}
\usepackage{subcaption}
\usepackage{xcolor}
\usepackage{graphicx}
\usepackage{graphics}
\usepackage{dsfont}
\usepackage[unicode]{hyperref}
\hypersetup{
    colorlinks=true,
    linkcolor=[rgb]{.8,.1,.1},
    citecolor=[rgb]{.22,.33,.88},
    urlcolor=[rgb]{.66,.11,.22}
}
\usepackage{amsmath,amsthm,amssymb,amsfonts,dsfont}
\usepackage[shortlabels]{enumitem} 
\usepackage[parfill]{parskip} 
\usepackage{pgfplots}
\pgfplotsset{compat=1.15}
\usepackage{mathrsfs}
\usetikzlibrary{arrows}

\def\R{\mathds R}
\def\C{\mathds C}
\def\d{\mathrm d}
\def\1{\mathds 1}
\def\wh{\widehat}
\def\wt{\widetilde}
\def\midd{\,\middle\vert\, }
\newcommand\restr[2]{{
\left.#1 \vphantom{\Bigr|}\right\lvert_{#2}
}}
\numberwithin{equation}{section}

\begin{document}
\sloppy
\binoppenalty=10000
\title{Inverse non-linear problem of the long wave run-up on coast\footnote{To appear in Journal of Ocean Engineering and Marine Energy}}
\author{Alexei Rybkin, Efim Pelinovsky, Oleksandr Bobrovnikov,   Noah Palmer,\\
Ekaterina Pniushkova and Daniel Abramowicz}

\date{November 2024}
\maketitle

{\scriptsize Alexei Rybkin,
Department of Mathematics and Statistics, University of Alaska
Fairbanks, PO Box 756660, Fairbanks, AK 99775.\\
\href{mailto:arybkin@alaska.edu}{arybkin@alaska.edu}

Efim Pelinovsky,
HSE University, Nizhny Novgorod, Russia
and
A.V. Gaponov-Grekhov Institute of Applied Physics, Nizhny Novgorod, Russia\\
\href{mailto:Pelinovsky@gmail.com}{Pelinovsky@gmail.com}

Oleksandr Bobrovnikov,
Department of Mathematics and Statistics, University of Alaska
Fairbanks, PO Box 756660, Fairbanks, AK 99775.\\
\href{mailto:obobrovnikov@alaska.edu}{obobrovnikov@alaska.edu}

Noah Palmer,
{Department of Applied Mathematics, University of Colorado Boulder, Boulder, CO.\\
\href{mailto:noah.palmer@colorado.edu}{noah.palmer@colorado.edu}}

Ekaterina Pniushkova,
Department of Biomedical Engineering, Northwestern University, Evanston IL.
\href{mailto:katya.pniushkova@gmail.com}{katya.pniushkova@gmail.com}

Daniel Abramowicz,
Department of Mathematics, University of San Francisco, San Francisco, CA.
\href{mailto:dlabramowicz@dons.usfca.edu}{dlabramowicz@dons.usfca.edu}

~}

\begin{abstract}
    {The study of the process of catastrophic tsunami-type waves on the coast makes it possible to determine the destructive force of waves on the coast. In hydrodynamics, the one-dimensional theory of the run-up of non-linear waves on a flat slope has gained great popularity, within which rigorous analytical results have been obtained in the class of non-breaking waves. In general, the result depends on the characteristics of the wave approaching (or generated on) the slope, which is usually not known in the measurements. Here we describe a rigorous method for recovering the initial displacement in a source localised in an inclined power-shaped channel from the characteristics of a moving shoreline. The method uses the generalised Carrier-Greenspan transformation, which allows one-dimensional non-linear shallow-water equations to be reduced to linear ones. The solution is found in terms of Erd\'elyi-Kober integral operator. Numerical verification of our results is presented for the cases of a parabolic bay and an infinite plane beach.}  
\end{abstract}
\section{Introduction}
With the devastating loss of life caused by tsunamis such as the Indian Ocean Tsunami in 2004 and the T\=ohoku Tsunami in 2011, predictive modelling of tsunami wave run-up is of great practical importance. In modern tsunami wave modelling, the shallow water, or long-wave, approximations are commonly used to predict inundation areas \cite{Levin16}. Such models require initial conditions to compute wave propagation. However, due to a lack of data on initial water displacement, models such as the Okada seismic model \cite{Okada92} are often used to generate the initial data. An alternative approach is to indirectly estimate  characteristics of the tsunami source through various inversion methods. Implementations have been used to recover the initial height of the water at the source \cite{Abe73}, the source location \cite{Fujii11} as well as fault motion \cite{Satake87} to name just three. The latter, through the inversion of data gathered for the wave signal, is crucial in such fields  as seismic hazard assessment. Through studying the accumulation of slip on each segment of a fault via the inverse problem, the prediction of earthquake recurrence intervals becomes increasingly more accurate. We suggest the recent review by K.~Satake \cite{Satake21} and the sources therein for more details on tsunami inversion methods, including waveform inversion, inverse modelling for the purpose of examining the tsunami source, and the generation of tsunami inverse refraction diagrams. Note that most methods assume that wave propagation is linear, while the tsunami wave run-up is a notoriously non-linear phenomenon. Unfortunately, inverse problems for non-linear PDEs are intractable in general.
{Thus, identifying a realistic class of bathymetries where non-linear inversion methods can be applied is a primary objective of our research.}

Currently, tsunami inundation calculations are conducted using numerical codes that model wave propagation from the source to the coast, validated against a series of benchmarks supported by experimental data. A significant part of this work involves analyzing the run-up of non-linear long waves on a flat slope, which has a rigorous analytical solution in the class of non-breaking waves using the Carrier-Greenspan transformation. The Carrier-Greenspan transformation simplifies the non-linear shallow-water equations to a linear wave equation with cylindrical symmetry (a particular class of the  Euler-Poisson-Darboux equation) \cite{Carrier58}. Within this framework, the run-up of waves generated on the slope is considered, using various forms of intial displacement with the initial displacement of the water surface, such as solitons \cite{Synolakis87}, Gaussian \cite{Carrier03} and Lorentz \cite{Pelin-Maz} pulses, $N$-waves \cite{Tadepalli}, cnoidal waves \cite{Synolakis88}, algebraic pulses \cite{Dobrokhotov}, and the Okada solution \cite{Tinti,Lovholt}. Naturally, the specific characteristics of the run-up depend on the features of the initial displacement at the source. Thus, attempts have been made to parametrise these formulas in order to reduce the number of initial perturbation parameters \cite{Didenkulova18,Lovholt}. More broadly, recent studies have also considered the fluid velocity at the source \cite{Kanoglu06,Rybkin19} and addressed boundary problems for waves approaching the coast \cite{Antuono07,Aydin20}. Similar approaches have also been developed for wave run-up in power-shaped bays \cite{Hartle21, Nicolsky18, Rybkin14,Rybkin21,Shimozono}.

In the papers cited above, the direct problem (the Cauchy problem) of the non-linear equations of the shallow water theory was solved. In this case,
since there are no measurements of wave parameters in the shelf zone, model functions were used as initial conditions.
In view of this, the inverse problem of recovering the initial conditions from the given (experimental or model) characteristics of the moving shoreline is of interest. This is especially important for fast estimates of tsunami waves in situations with uncertain wave properties during real events.

In this work we consider the problem of recovering the shape of an incident wave from the known oscillations of a moving shoreline. This problem was first considered in \cite{Rybkin23} and our work here is a generalisation of those results to a more diverse set of bathymetries. In this case, the following restrictions are imposed: the tsunami source is located on a slope at an arbitrary distance from the shoreline. Two configurations of the bottom relief are considered: a flat slope and an inclined parabolic channel. The solution of the inverse problem is found using the Abel transform in the class of non-breaking waves.

Our work here is organised as follows. In Section \ref{sec:SWE}, we introduce the shallow water framework our model is built upon. In Section \ref{sec:dir_inv_prob} we give the statement of both the direct and inverse problem and introduce the Carrier-Greenspan hodograph transformation on which our method is based. We solve both the direct and inverse problems in Section \ref{sec:shoreline} through the derivation of what we call the shoreline equation, an equation relating the mechanical energy of the wave at the shoreline and the initial wave profile. Section \ref{sec:supp_est} discusses the recovery of certain characteristics of the initial wave. In Section \ref{sec:num} we give numerical verifications of our method. Finally, in Section \ref{sec:disc} we give some concluding remarks and discuss some potential future directions.

\section{Shallow water equations (SWE)}\label{sec:SWE}

\begin{figure}[h]
    \begin{subfigure}{.45\linewidth}
        \definecolor{wwwwww}{rgb}{1,0.95,0.45}
        \definecolor{qqqqff}{rgb}{0.,0.45,95.}
        \definecolor{qqqqffr}{rgb}{0.9,0.3,0.3}

      \caption{$(x, z)$ cross-sectional view;}\label{fig:xoz}
    \end{subfigure}\hfill
      \begin{subfigure}{.45\linewidth}
  \begin{center}
    \definecolor{qqqqff}{rgb}{0.,0.45.,95.}
\definecolor{qqqqffr}{rgb}{0.9,0.3,0.3}
\definecolor{wqwqwq}{rgb}{1,0.95,0.45}
%
           \caption{$(y, z)$ cross-sectional view;}\label{fig:xoy}
        \end{center}
    \end{subfigure}
    \begin{subfigure}{\linewidth}
     \begin{center}
      \includegraphics[width = \linewidth]{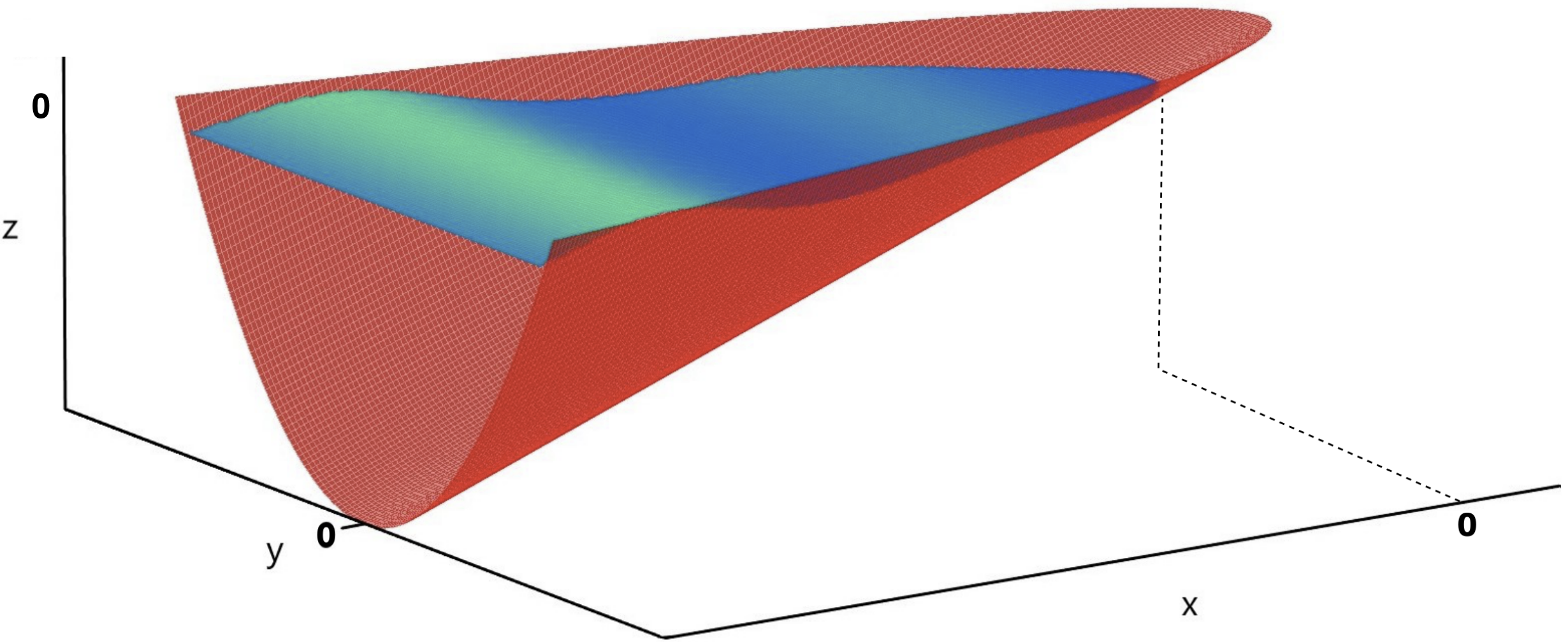}
     \end{center}
      \caption{3D view;}\label{3dview}
    \end{subfigure}
    \caption{Geometrical representations of a power-shaped bathymetry resembling the case $m=2$. In \ref{fig:xoz} we have cross sectional view of the $xOz$ plane, in \ref{fig:xoy} a cross sectional view of the $yOz$ plane, and in \ref{3dview} a 3-dimensional view of the bay and an $N$-wave.}\label{finalfig}
\end{figure}

The shallow water equations (SWE) are a set of non-linear, hyperbolic PDEs which are commonly used to model tsunami wave run-up. The 2+1 (that is the unknown functions are of two spatial variables \(x,y\) and the temporal variable \(t\)) SWE are a simplification of the Euler equations, a highly non-linear 3+1 (that is the unknown functions are of three spatial variables \(x,y,z\) and the temporal variable \(t\)) system. They can be derived with the truncation of Taylor expansions of non-linear terms and the assumptions of no vorticity, small vertical velocity, and small depth/wavelength and wave height/depth ratios.
The 2+1 SWE can be further reduced to a 1+1 system (that is now the functions are of \(x\) and \(t\) only) by assuming that the bathymetry is centred along the $x$ axis and is uniformly inclined. For the bathymetries we are concerned with here (see Fig. \ref{finalfig}), power-shaped bays with $y$ cross-section $|y|^m$, the non-linear SWE in non-dimensional units are given as
\begin{equation}\label{SWE}
\begin{split}
      \partial_t\eta+u\left(1+\partial_x\eta\right)+\frac{m}{m+1}\left(x+\eta\right)\partial_x u&=0,  \\
      \partial_t u+u\partial_x u+\partial_x \eta&=0, 
\end{split}
\end{equation}
where $u(x, t)$ is the depth averaged flow velocity over the corresponding cross-section, and $\eta(x, t)$ is the water displacement exceeding the unperturbed water level.
The total perturbed water depth is given as $H(x,t)=h(x) + \eta(x,t)$ along the $x$ axis, where $h(x)$ is the depth of the bay, and so in dimensionless units we simply have $h(x) = x$. Typically, the system seen in \eqref{SWE} is given in dimensional units. The substitution
\begin{equation}\label{eq:dim}
\wt{x}=(H_0/\alpha){x},\quad \wt{t}=\sqrt{H_0/g}\,{t}/\alpha,\quad \wt{\eta}=H_0{\eta},\quad \wt{u}=\sqrt{H_0 g}~{u},
\end{equation}
where $H_0$ is the characteristic height of the wave, $\alpha$ is the slope of the bathymetry and $g$ is the acceleration of gravity, turns the dimensionless system into one with dimension (dimensional variables are the ones with tildes). The shoreline in the physical plane (i.e., the wet/dry boundary) is given by
\begin{equation}\label{physicalshore}
x+\eta(x,t)=0.
\end{equation}
The solution to \eqref{physicalshore}, let us call it $x_0(t)$, describes the run-up and draw-down of the tsunami wave. We consider the initial value problem for \eqref{SWE} with
typical initial conditions characterised the instantaneous bottom displacement (see for instance \cite{Okada85})
\begin{equation}\label{SWE IC}
\begin{aligned}
\eta(x,0) = \eta_0(x), \\
u(x,0) = 0.
\end{aligned}
\end{equation}
While the choice of zero initial velocity may be restrictive in physical application there are good reasons for this choice. For earthquake generated tsunamis it is typical to assume that the initial velocity is zero. Additionally, it is a convenient choice, a common technique in finding solutions to the SWE

\section{Statement of the direct and inverse problems}\label{sec:dir_inv_prob}
In this paper we investigate the following direct problem: knowing the initial displacement of the water $\eta_0(x)$ and assuming zero initial velocity of the water, we find the movement of the shoreline $x_0(t)$.
The direct problem was solved both numerically and analytically by many authors (see the references in the introduction).
The corresponding inverse problem then consists of restoring the initial displacement of the water, assuming zero initial velocity and knowing the shoreline movement $x_0(t)$ and the time of an earthquake, i.e., zero time.
It is worth noting that a non-linear inverse problem, as is the case with our problem, presents several challenges in terms of deriving a solution.
The main difficulty is that the shoreline is moving.
The Carrier-Greenspan transform allows to reduce the original problem to a linear one on $\R_{>0}$.
\subsection*{Carrier-Greenspan Transform}
The Carrier-Greenspan (CG) hodograph transform, introduced in \cite{Carrier58}, can be used to linearise \eqref{SWE} into a form which can then be solved using Hankel transforms \cite{Hilbert}. We use the form of the CG transform, originally introduced for power-shaped bays in \cite{Tuck72}:
\begin{equation}\label{CG}
    \varphi(\tau,\sigma) = u(x,t),\quad \sigma = x + \eta(x,t),\quad \psi(\tau,\sigma) = \eta(x,t) + u^2(x,t)/2,\quad \tau = t-u(x,t).
\end{equation}
Applying \eqref{CG} to \eqref{SWE} yields the {linear hyperbolic system}
\begin{equation}
\begin{split}
\partial_\tau \psi + \frac{m}{m+1} \sigma \partial_\sigma\varphi + \varphi &= 0,\\
\partial_\tau\varphi + \partial_\sigma \psi &= 0,
\end{split}
\end{equation}
which is often written as the second order equation
\begin{equation} \label{linearsys}
\partial^2_\tau\psi = \frac{m}{m+1}\sigma\partial^2_\sigma\psi + \partial_\sigma\psi.
\end{equation}

We therefore obtain a linear hyperbolic equation \eqref{linearsys} from a non-linear system \eqref{SWE}. Physically, $\sigma$ denotes wave height from the bottom, $\tau$ is a delayed time, $\varphi$ is the flow velocity, and $\psi$ can be called the total
energy.
The CG transforms main benefit is that the moving shoreline $x_0(t)$ is fixed at $\sigma = 0$. Nevertheless, the CG transform has some notable drawbacks; for one, the ICs become complicated in the hodograph coordinates, making standard techniques difficult to apply.  However, by setting the initial velocity of the wave to be zero, that is $u_0(x)=0$, one avoids this issue. While, this premiss is restrictive, it is typical when considering earthquake generated tsunamis.
Thus, we assume this condition which is equivalent to $\varphi(0,\sigma)=0$, and so \eqref{SWE IC} becomes

\begin{equation} \label{transformedIC}
\restr{\varphi}{\Gamma} = 0,\quad \restr{\psi}{\Gamma} = \psi_0(\sigma)= \eta_0(\gamma(\sigma))
\end{equation}
where $\Gamma$ is the vertical line $(0,\sigma)$ in the hodograph plane and $x=\gamma(\sigma)$ solves $\sigma= \eta(x,t)+x$. Additionally, the regular singularity at $\sigma=0$ causes computational difficulties at the shoreline. Finally, we note that the transformation only works provided it is invertible, i.e. the wave does not break \cite{Rybkin21}, so we must surmise this going forward. 

\section{The Shoreline Equation}\label{sec:shoreline}

\def\F{\mathcal F}
\def\H{\mathcal H}

In this section we derive what we call the shoreline equation of an arbitrary power-shaped bay. Specifically, we derive an equation relating $\psi(\tau,0)$, the energy of the water at the shoreline, and $\psi(0,\sigma)=\psi_0(\sigma)$ the initial displacement of the water. Notably, the direct problem has previously been solved for power-shaped bathymetries (see, for instance, \cite{Garashin16},\cite{Didenkulova11}) and the inverse problem in the narrow case of a plane beach \cite{Rybkin23}. Here the direct problem is solved both analytically, as follows in this section, and numerically, as can be seen in Section 6, to ensure that propagation of the wave is being accounted for as described in \cite{Satake87}. Since the energy at the shoreline can be computed from the movement of the shoreline $x_0(t)$, the shoreline equation allows us to easily solve the inverse problem and recover $\eta_0(x)$ after converting back into physical space.

We start with the bounded analytical solution to the initial value problem ({\ref{linearsys},~\ref{transformedIC}}), which is given in \cite{Rybkin21}:
\begin{equation}\label{hankelpsi}
    \psi(\tau,\sigma)=\sigma^{-\frac{1}{2m}}\int_0^\infty 2k\left(\int_0^\infty \psi_0(s) s^{\frac{1}{2m}}J_{\frac{1}{m}}(2k\sqrt{s})~\d s\right)\cos\left(\sqrt{\frac{m}{m+1}}k\tau\right) J_{\frac{1}{m}}(2k\sqrt{\sigma})~\d k,
\end{equation}
where $J_{\frac{1}{m}}$ is the Bessel function of the first kind of order $1/m$ and $\Gamma(z)$ is the gamma function.
Since $J_{\nu}(z) = z^\nu 2^{-\nu}/\Gamma\left( \nu + 1 \right) + {o}(1)$ as $z\to +0$, we obtain
\begin{equation}
    \psi(\tau,0)=\frac{2}{\Gamma(\frac{1}{m}+1)}\int_0^\infty k^{\frac{1}{m}+1}\left(\int_0^\infty \psi_0(s)s^{\frac{1}{2m}}J_{\frac{1}{m}}(2k\sqrt{s})~\d s\right)\cos\left(\sqrt{\frac{m}{m+1}}k\tau\right) ~\d k,
\end{equation}
which after the substitution $\lambda=\sqrt{s} $ and $\wh{\psi}_0(\lambda)=\psi_0(\lambda^2)$ becomes
\begin{equation}\label{psi2}
    \psi(\tau,0)=\frac{4}{\Gamma(\frac{1}{m}+1)}\int_0^\infty k^{\frac{1}{m}+1}\left(\int_0^\infty \wh{\psi}_0(\lambda)\lambda^{\frac{1}{m}+1}J_{\frac{1}{m}}(2k\lambda)~\d \lambda\right)\cos\left(\sqrt{\frac{m}{m+1}}k\tau\right) ~\d k.
\end{equation}
Now, define the modified Hankel transform as
\begin{equation}\label{mhankel}
\left[\wh{\H}_\nu f(r)\right]   (k)=k^{-\nu}\int_0^\infty f(r)J_\nu(kr)r^{\nu+1}~\d r.
\end{equation}

Note that
\begin{equation}
    \left[ \wh\H_\nu f(r) \right](\lambda) = \lambda^{-\nu} \left[ \H_\nu r^\nu f(r) \right](\lambda),
\end{equation}
where $\H_\nu$ is the standard Hankel transform, and so we observe that $\wh \H_\nu$ is self-inverse.

So, applying \eqref{mhankel} to \eqref{psi2} we have
\begin{equation}\label{psi3}
    \psi(\tau,0)=\frac{2^{2+\frac{1}{m}}}{\Gamma(\frac{1}{m}+1)}\int_0^\infty
    k^{1+\frac{2}{m}}\left[\wh{\H}_{\frac{1}{m}} \wh\psi_0(\lambda)\right](2k)
    \cos\left(2\pi \xi k\right)~\d k,
\end{equation}
where $\xi=(\sqrt{m/(m+1)}\tau)/2\pi = : q^{-1} \tau$. Let $g(k)= k^{1+\frac{1}{2m}}
\left[\wh{\H}_{\frac{1}{m}} \wh \psi_0(\lambda)\right](2k)$ and denote
\begin{equation}\label{Fouriercosine}
\left[\F_c f(t)\right](\xi)=\int_0^\infty f(t)\cos(2\pi\xi t)~\d t,
\end{equation}
as the Fourier cosine transform. Then we obtain
\begin{equation}\label{almostthere}
        \psi(\tau,0)=\frac{2^{2+\frac{1}{m}}}{\Gamma(\frac{1}{m}+1)}\int_0^\infty g(k)\cos(2\pi\xi k)~\d k
        =\frac{2^{2+\frac{1}{m}}}{\Gamma(\frac{1}{m}+1)}\left[\F_c g(k)\right](\xi) =
        \frac{2^{2+\frac{1}{m}}}{\Gamma(\frac{1}{m}+1)}\left[\F_c k^{1+\frac{1}{2m}}
        \left[\wh{\H}_{\frac{1}{m}} \wh \psi_0(\lambda)\right](2k)\right](\xi) 
        .
\end{equation}

So \eqref{almostthere} allows us to solve the direct problem. For that one would need to find $\psi_0(\sigma)$ from the Carrier-Greenspan transform as
\begin{equation}
\eta_0\left(x\right)=\psi_0\left(\sigma\right),\quad \sigma={x+\eta_0\left(x\right)},
\end{equation}

then compute two integral transforms, and finally return to the $(x,t)$ space using the inverse CG transform, which at the shoreline becomes
\begin{equation}
\psi\left(\tau,0\right)=-x_0\left(t\right)+ \dot x_0\left(t\right)^2/2,\quad \tau = t - \dot x_0\left(t\right).
\end{equation}

\subsection*{Solution to the Inverse problem}
In this section we invert the transform given in \eqref{almostthere} in order to solve the inverse problem.
 
Applying the inverse Fourier cosine transform to \eqref{almostthere} we obtain
\begin{equation}
    \begin{aligned}
         \frac{\Gamma(\frac{1}{m}+1)}{2^{2+\frac{1}{m}}}\psi(q\xi,0)&= \left[\F_c g(k)\right](\xi)\\
         \frac{\Gamma(\frac{1}{m}+1)}{2^{2+\frac{1}{m}}}\left[\F_c^{-1}\psi(q\xi,0)\right](k)&= k^{1+\frac{1}{2m}}\left[\wh{\H}_{\frac{1}{m}} \wh \psi_0(\lambda)\right](2k)\\
    \frac{\Gamma(\frac{1}{m}+1)}{2^{2+\frac{1}{m}}k^{1+\frac{1}{2m}}}\left[\F_c^{-1}\psi(q\xi,0)\right](k)&=\left[\wh{\H}_{\frac{1}{m}} \wh \psi_0(\lambda)\right](2k).
    \end{aligned}
    \end{equation}

Applying the inverse Hankel transform and utilising the identity \cite{grad-ryzh}
\begin{equation}
    J_\nu(z) = \frac{2\left( \frac z 2  \right)^\nu }{\Gamma\left( \nu+\frac 1 2  \right)\sqrt \pi }\int_{0}^{1}\cos zs (1-s^2)^{\nu-\frac 1 2}~\d s \quad\text{for }z>0,~\nu>-\frac 1 2,
\end{equation}
we obtain
\begin{equation}
\begin{aligned}
     \wh \psi_0 (\lambda) &=
     2 \Gamma(1 + 1 / m) \lambda^{- \frac{1}{m}}\int_0^\infty \psi(q \xi, 0) \cos (2 \pi k \xi)\,\d \xi J_{\frac{1}{m}}(2 k \lambda)\,\d (2k)
     \\ & \quad\text{[use integral representation of Bessel function]}
     \\ &= 2 \Gamma\left(1 + \frac{1}{m}\right) \lambda^{- \frac{1}{m}}\int_0^\infty k^{- \frac{1}{m}} \int_0^\infty \psi(q \xi, 0) \cos(2 \pi k \xi )\,\d \xi \frac{2 (k \lambda)^{\frac{1}{m}}}{\Gamma\left(\frac{1}{m} + \frac{1}{2}\right)\sqrt{\pi}}\int_0^1 \cos(2 k \lambda s) (1-s^2)^{\frac{1}{m}-\frac{1}{2}}\,\d s\,\d(2k)
     \\ & \quad\text{[regroup and change order of integration]}
     \\ & =\frac{8 \Gamma(1 + 1 / m)}{\Gamma(1 / m + 1 / 2)\sqrt{\pi}} \int_0^1 (1-s^2)^{\frac{1}{m}-\frac{1}{2}} \int_0^\infty \int_0^\infty \cos(2 \pi k \xi) \cos(2 k \lambda s) \psi(q \xi, 0)\,\d \xi\,\d k \,d s.
\end{aligned}
\end{equation}
Further change of variables \(r = \lambda s / \pi\)   
yields
\begin{equation}
\begin{aligned} 
    \wh \psi_0 (\lambda) &=\frac{8 \Gamma(1 + 1 / m)}{\Gamma(1 / m + 1 / 2)\sqrt{\pi}} \int_0^{{\lambda}/{\pi}} \left(1- \frac{\pi^2 r^2}{\lambda^2}\right)^{\frac{1}{m}-\frac{1}{2}} \int_0^\infty \int_0^\infty \cos(2 \pi k \xi) \cos(2 \pi k r) \psi(q \xi, 0)\frac{\pi}{\lambda}\,\d \xi\,\d k \,\d r 
    \\ & \quad\text{[split }8 = 2\cdot 4 \text{ and regroup]}
    \\ &=\frac{2\sqrt{\pi} \Gamma(1 + 1 / m)}{\lambda\Gamma(1 / m + 1 / 2)} \int_0^{{\lambda}/{\pi}} \left(1- \frac{\pi^2 r^2}{\lambda^2}\right)^{\frac{1}{m}-\frac{1}{2}} 4\int_0^\infty \int_0^\infty \cos(2 \pi k \xi) \cos(2 \pi k r) \psi(q \xi, 0)\,\d \xi\,\d k \,\d r
    \\ & \quad\text{[Fourier transforms cancel]}
    \\ &=
    \frac{2\sqrt\pi \Gamma(1+1/m)}{\lambda \Gamma(1/m+1/2)}\int_0^{\lambda/\pi}\left( 1 - \left( \frac{\pi r}{\lambda} \right)^2\right)^{1/m-1/2} \psi(q r,0)~\d r.
\end{aligned}
\end{equation}
Upon switching back to variable $\sigma$ one obtains (note that above \(r\) corresponds to \(\xi\))
\begin{equation}\label{shoreline}
    \psi_0(\sigma^2) = \frac{2\sqrt \pi\Gamma(1+1/m)}{\sigma^2\Gamma(1/m+1/2)}\int_0^{\sigma^2/\pi}\left( 1-\left( \frac{\pi\xi}{\sigma^2} \right)^2 \right)^{1/m-1/2}\psi(q \xi,0)~\d\xi.
\end{equation}
Now denote
\begin{equation}\label{Abelforward}
\left[\mathcal{A}_{\alpha}f(x)\right](s)=\int_0^s\frac{f(x)~\d x}{(s^2-x^2)^\alpha}
\end{equation}
as the singular Abel type integral of order $\alpha$ as seen in \cite{Deans}. After a straightforward substitution one obtains
\begin{equation}\label{TheShorelineEquation}
\psi_0(\sigma^2) = \frac{2 \Gamma\left( \frac{m+1}{m } \right)}{\Gamma\left( \frac 1 m +\frac 1 2 \right)\sqrt \pi}\sigma^{-\frac 4 m}\left[ \mathcal A_{1/2-1/m} \psi\left( \frac{q r}{\pi},0 \right) \right](\sigma^2).
\end{equation}
Now we can use \eqref{TheShorelineEquation} to solve the inverse problem as follows: from the shoreline movement $x_0(t)$ we find ${\psi(\tau,0) = -x_0(t) + \dot x_0(t)^2/ 2}$ and $\tau = t - \dot x_0(t)$, after that from \eqref{TheShorelineEquation} we find $\psi_0(\sigma)$, and finally we find ${\eta_0(x) = \psi_0(\sigma)}$ and ${x = \sigma - \eta_0(x)}$.
Note that the algorithm laid out above uses dimensionless units. The substitution \eqref{eq:dim} should be used when dealing with dimensional measurements.

\subsection*{Some Remarks}

The transform defined in \eqref{Abelforward} is in fact Erd\'elyi-Kober fractional integration operator (see for example \cite{Sneddon}). This operator is closely connected \cite{Erdelyi70} to the Euler-Poisson-Darboux (EPD) equation, which one can obtain from SWE \eqref{SWE} by taking in the CG transform \eqref{CG} $\sigma^2 = x+\eta(x,t)$. Moreover, one can use the CG transform used in \cite{Garashin16} to obtain the IBVP for the EPD equation, that solves the inverse problem, and after that use the technique laid out in \cite{Erdelyi70} to solve it.  The only disadvantage of this approach is that it only applies for $m>2$, while our method works for any positive $m$.
In \cite{Erdelyi70} there is the claim, that this restriction can be relaxed to any positive $m$, however we have not investigated that.

It is worth noting that \eqref{Abelforward} has an inverse formula for $\alpha\in(0,1)$ \cite{Deans}, given as
\begin{equation}
\left[\mathcal{A}_\alpha^{-1}f(x)\right](s)=\frac{2\sin(\alpha\pi)}{\pi}\frac{\d}{\d s}\int_0^s\frac{xf(x)~\d x}{(s^2-x^2)^{1-\alpha}}.
\end{equation}
Thus, for all $m>2$ we can invert \eqref{TheShorelineEquation} to obtain (after substituting  $s=\sigma^2$)
\begin{equation}\label{fur-tichmarsh}
\psi(qr/\pi,0) = \frac{\sqrt{\pi}}{\Gamma(1+\frac{1}{m})\Gamma(\frac{1}{2}-\frac{1}{m})}\frac{\d}{\d r}\left[\mathcal{A}_{\frac{m+2}{2m}}s^{\frac{2}{m}}\psi_0(s)\right](r).
\end{equation}
This allows to solve the direct problem using one integral operator, rather then composing two Fourier transform for $m>2$.

\subsection*{Particular cases}

In the two most interesting cases, that is the case of the infinite plane beach corresponding to $m=\infty$ and a parabolic bay for $m=2$, our solution can be shown to reduce down to particularly nice forms. For $m=\infty$ we have $q = 2\pi$\, and so \eqref{TheShorelineEquation} easily simplifies to 
\begin{equation} \label{m=inf}
\psi_{0}(\sigma^2) = \frac{2}{\pi}\left[\mathcal{A}_{\frac{1}{2}}\psi(2r, 0)\right](\sigma^2).
\end{equation}
{The substitution $\sigma'= \sigma^2$ and $\tau' = \tau/2$ turns \eqref{m=inf} to the form obtained in \cite{Rybkin23}.}

For $m=2$
we have $q = \pi\sqrt 6$, and so \eqref{TheShorelineEquation} simplifies to

\begin{equation}\label{m=2}
\psi_0(\sigma^2) = \frac{1}{\sigma^2}\left[\mathcal{A}_0\psi(\sqrt{6}r, 0)\right](\sigma^2) 
=\frac{1}{\sigma^2}\int_0^{\sigma^2}\psi(\sqrt{6}r,0)~\d r.
\end{equation}

\section{Estimate of the shape of the incoming wave}\label{sec:supp_est}
In this section we give the exact lower bound for the support of the initial water displacement in terms of the shoreline data. 
First we remind that for a scalar-valued function $f:X\to \C$ its support is the set $\operatorname{supp} f = \left\{ x\in X\midd f(x)\neq 0 \right\}$. 
For the case $m = 2$ we have
\begin{equation}\label{m2Simplified}
    \psi_0(\sigma^2) = \sigma^{-2}\int_0^{\sigma^2} \psi(\sqrt 6 r,0)~\d r,
\end{equation}
and so we deduce that
$
    \inf\operatorname{supp} \psi_0(\sigma^2) =  \inf\operatorname{supp}\psi\left(\sqrt 6 r,0\right)
$.

For $m>2$ we can 
use Titchmarsh's convolution  theorem, which states (see \cite{Tichmarsh} for details) that if
\begin{equation}
  \restr{(f*g)}{x\in (0,a)} = \restr{\left( \int_0^x f(t)g(x-t)~\d t \right)}{x\in(0,a)} = 0,
\end{equation}
and $g(x)>0$ on $(0,a)$, then $f(x) = 0$ almost everywhere on $(0,a)$.
From \eqref{fur-tichmarsh} we have
\begin{equation}
  \psi\left( \frac{qr}{\pi},0  \right) = C(m) \frac{\d}{\d r}
  \left( \int_0^r \left( \psi_0(s)s^{\frac 2 m} (r+s)^{\frac{2+m}{2m}}  \right)(r-s)^{\frac{2+m}{2m}}
  ~\d s\right)
  ,
\end{equation}
and so we deduce that $\inf\operatorname{supp} \psi_0(s) \leq \inf\operatorname{supp}\psi\left( \frac{qr}{\pi},0  \right)$.
The inverse inequality
immediately follows from \eqref{TheShorelineEquation},
and so combining these results we obtain $\inf\operatorname{supp} \psi_0(s) = \inf\operatorname{supp}\psi\left( \frac{qr}{\pi} ,0 \right)$.

So we can express the lower bound of the support of $\psi_0(\sigma)$. Since $\psi_0(\sigma) = \eta_0(x)$ and $\sigma = x + \eta_0(x)$, we can obtain the exact lower bound of the support $\eta_0(x)$, namely $\inf\operatorname{supp} \eta_0(x) = \inf \operatorname{supp} \psi_0(\sigma)$. In simple language that means that we can express how far from the shore the displacement is at the time of an earthquake.

\section{Numerical Computations}\label{sec:num}

\begin{figure}[p]
    \centering
    \centerline{\textbf{Initial Wave Profiles}}
  
    \begin{subfigure}{0.45\textwidth}
    \centering
        \includegraphics[width=\linewidth]{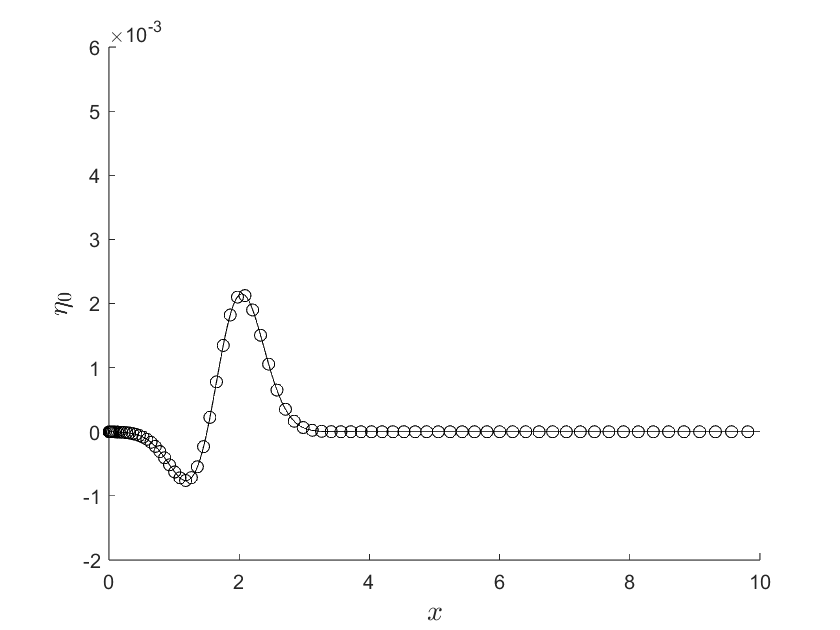}
        \caption{$m=1$}
        \label{m1_nwave}
    \end{subfigure}%
    \begin{subfigure}{0.45\textwidth}
    \centering
        \includegraphics[width=\linewidth]{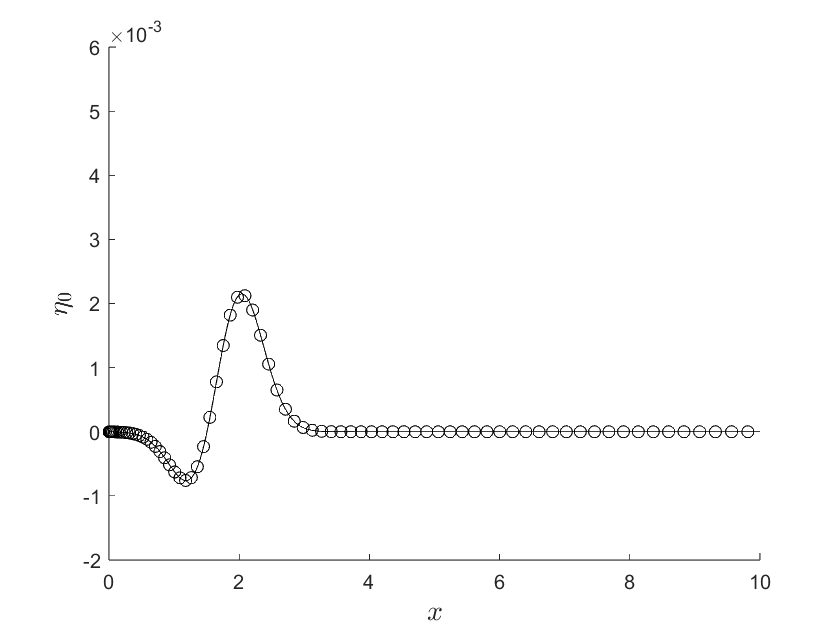}
        \caption{$m=2$}
        \label{m2_nwave}
    \end{subfigure}
   
    \begin{subfigure}{0.45\textwidth}
    \centering
        \includegraphics[width=\linewidth]{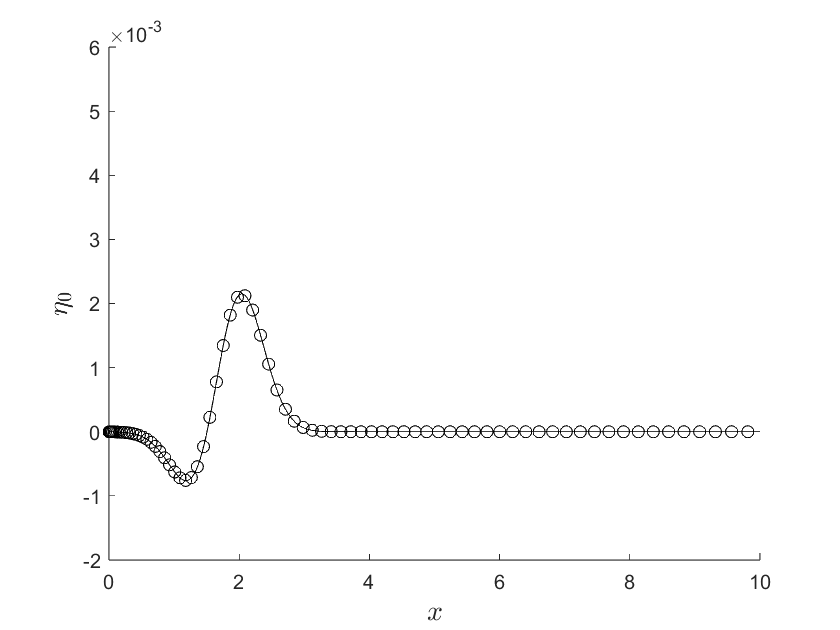}
        \caption{$m=3$}
        \label{m3_nwave}
    \end{subfigure}%
    \begin{subfigure}{0.45\textwidth}
    \centering
        \includegraphics[width=\linewidth]{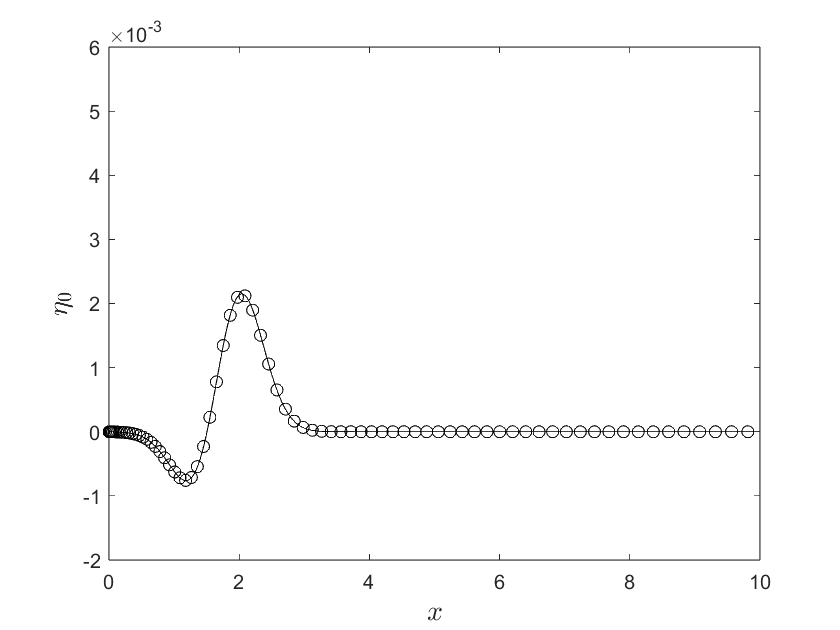}
        \caption{$m=\infty$}
        \label{minf_nwave}
    \end{subfigure}%
    \caption{A comparison of an initial displacement of an $N$-wave with the displacement predicted by our model for varying power-shaped bays. The solid black line gives the exact initial displacement and the open circles denote the initial displacement predicted by our model.}
    \label{numerical_nwave}
    \end{figure}

\begin{figure}[p]
    \centering
    \centerline{\textbf{Vertical Shift Profiles}}
  
    \begin{subfigure}{0.45\textwidth}
    \centering
        \includegraphics[width=\linewidth]{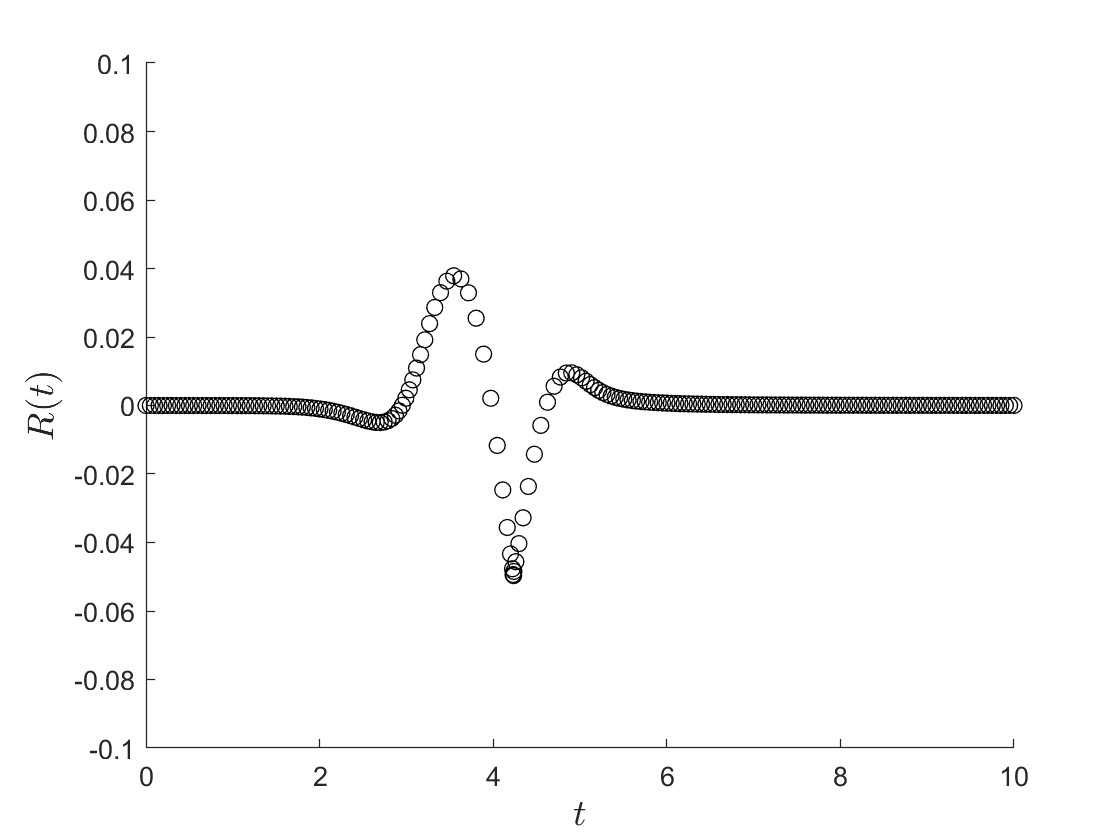}
        \caption{$m=1$}
        \label{m1_nvert}
    \end{subfigure}%
    \begin{subfigure}{0.45\textwidth}
    \centering
        \includegraphics[width=\linewidth]{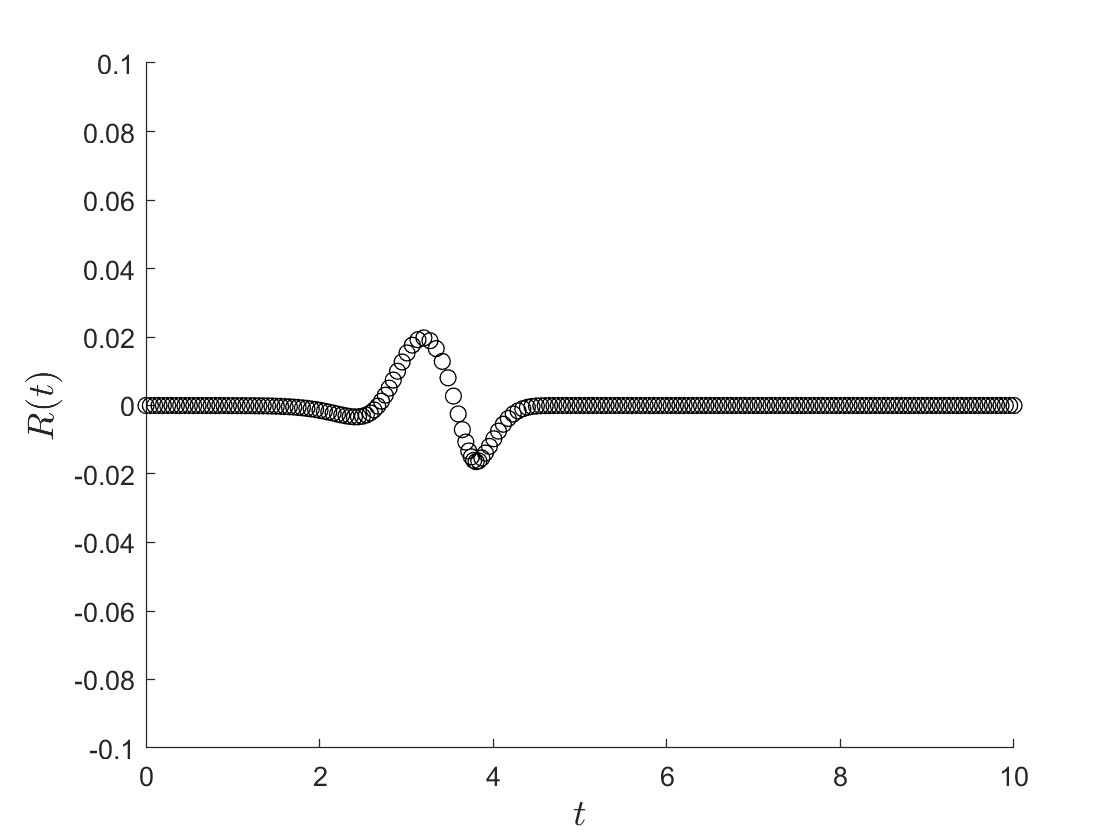}
        \caption{$m=2$}
        \label{m2_nvert}
    \end{subfigure}
   
    \begin{subfigure}{0.45\textwidth}
    \centering
        \includegraphics[width=\linewidth]{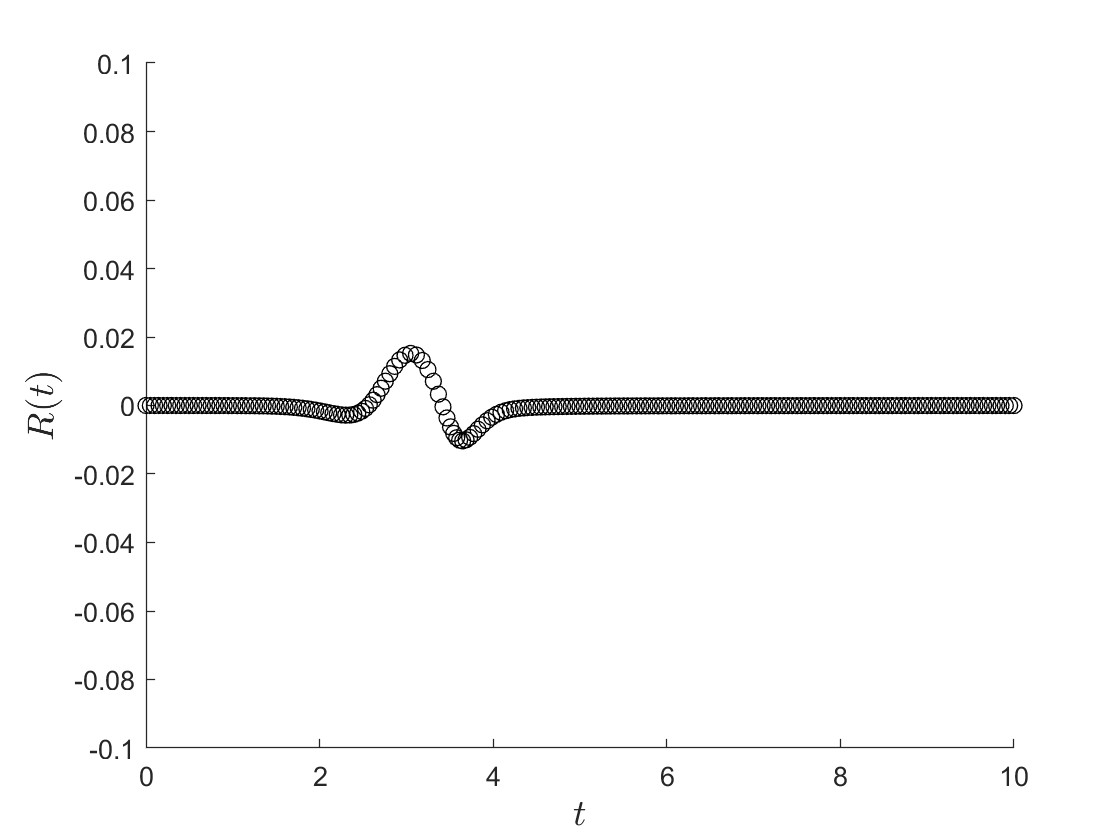}
        \caption{$m=3$}
        \label{m3_nvert}
    \end{subfigure}%
    \begin{subfigure}{0.45\textwidth}
    \centering
        \includegraphics[width=\linewidth]{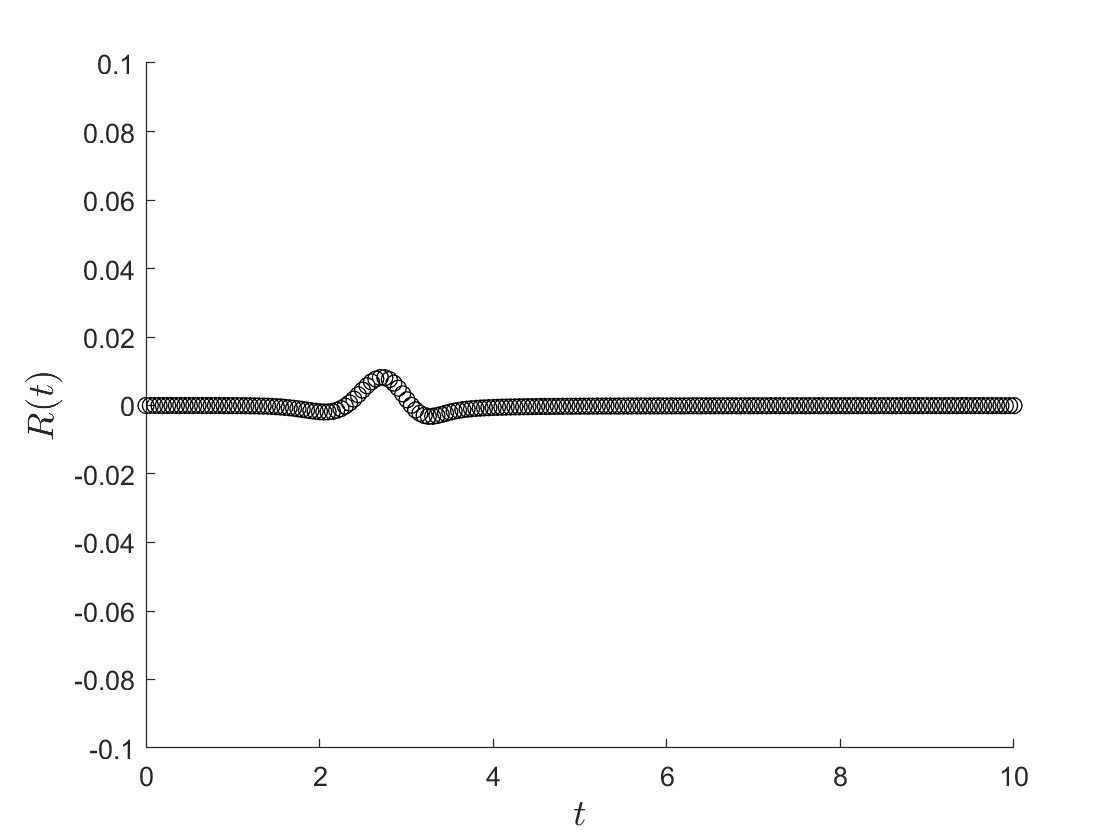}
        \caption{$m=\infty$}
        \label{minf_nvert}
    \end{subfigure}%
    \caption{Estimated vertical shift for an initial $N$-wave displacement corresponding to various bathymetries, where $R(t) = -x_0(t)/\alpha$, where $\alpha = 1$.}
    \label{shorelinefornwave}
    \end{figure}

\begin{figure}[p]
        \centering
        \centerline{\textbf{Initial Wave Profiles}}
        
        \begin{subfigure}{0.45\textwidth}
        \centering
            \includegraphics[width=\linewidth]{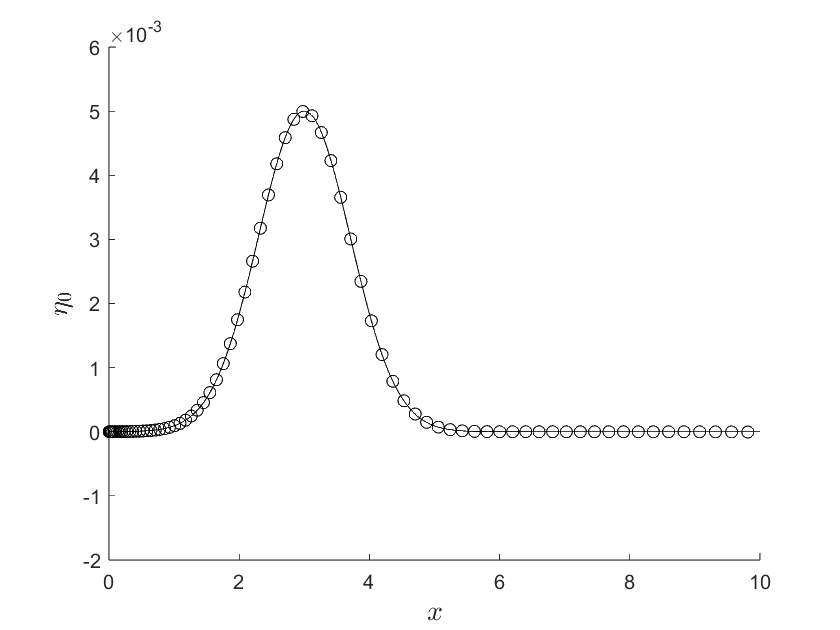}
            \caption{$m=1$}
            \label{m1_gaussian}
        \end{subfigure}%
        \begin{subfigure}{0.45\textwidth}
        \centering
            \includegraphics[width=\linewidth]{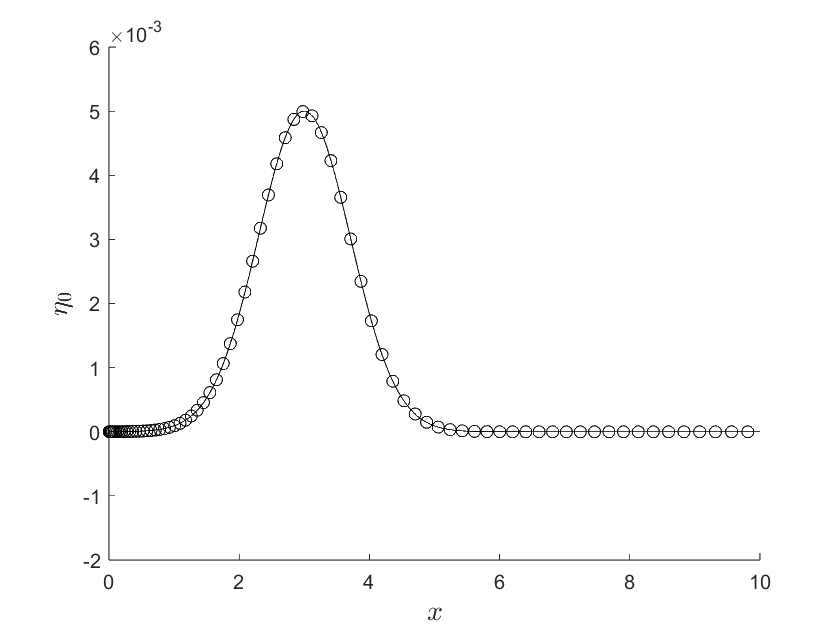}
            \caption{$m=2$}
            \label{m2_gaussian}
        \end{subfigure}
        
        \begin{subfigure}{0.45\textwidth}
        \centering
            \includegraphics[width=\linewidth]{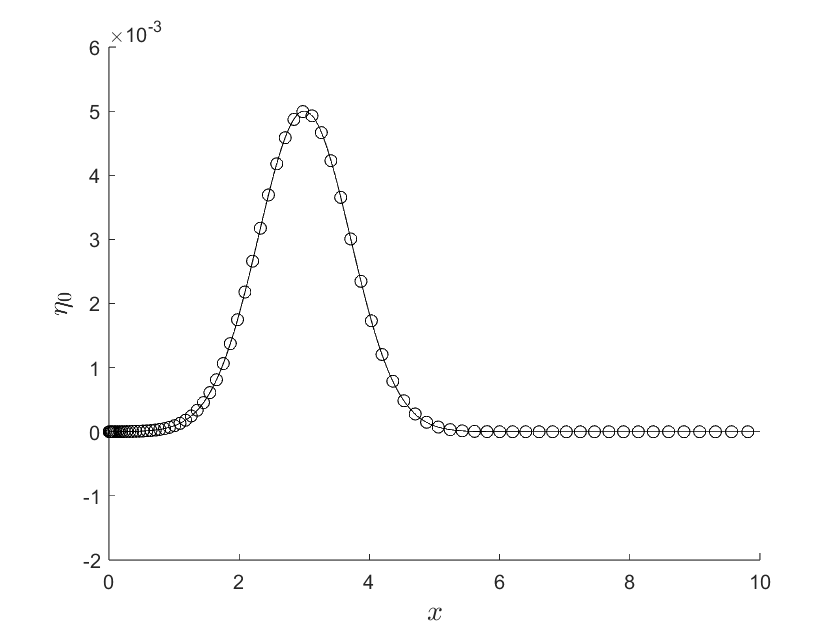}
            \caption{$m=3$}
            \label{m3_gaussian}
        \end{subfigure}%
        \begin{subfigure}{0.45\textwidth}
        \centering
            \includegraphics[width=\linewidth]{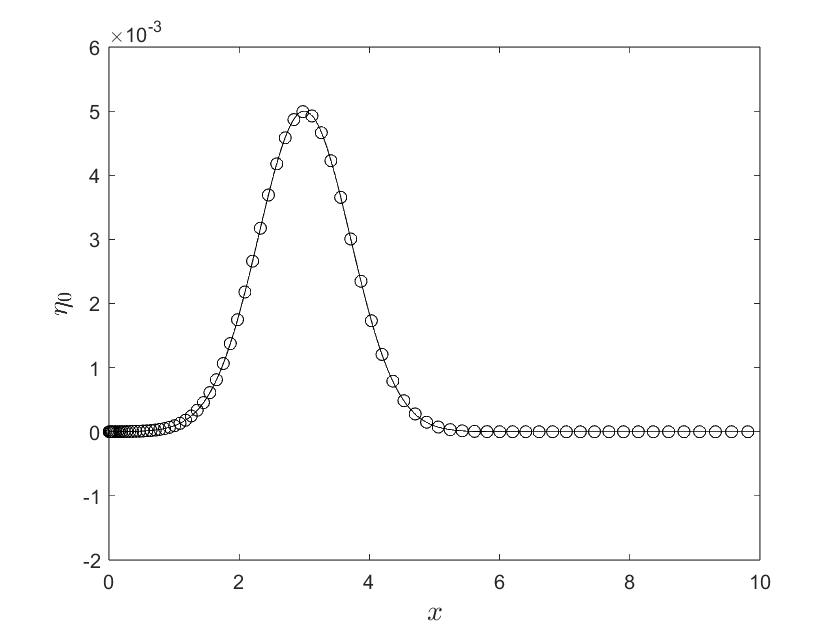}
            \caption{$m=\infty$}
            \label{minf_gaussian}
        \end{subfigure}%
        \caption{A comparison of an initial displacement of an Gaussian wave with the displacement predicted by our model for varying power-shaped bays. The solid black line gives the exact initial displacement and the open circles denote the initial displacement predicted by our model.}
        \label{numerical_gaussian}
        \end{figure}

\begin{figure}[p]
        \centering
        \centerline{\textbf{Vertical Shift Profiles}}
      
        \begin{subfigure}{0.45\textwidth}
        \centering
            \includegraphics[width=\linewidth]{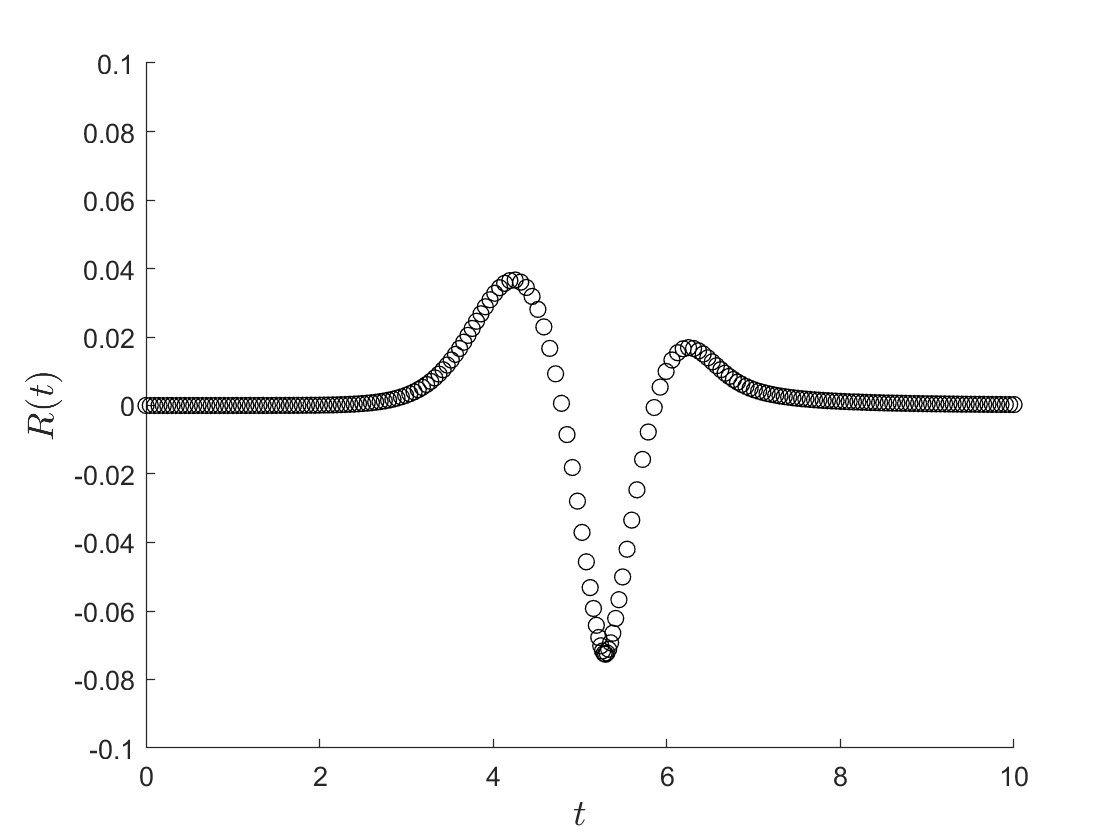}
            \caption{$m=1$}
            \label{m1_gvert}
        \end{subfigure}%
        \begin{subfigure}{0.45\textwidth}
        \centering
            \includegraphics[width=\linewidth]{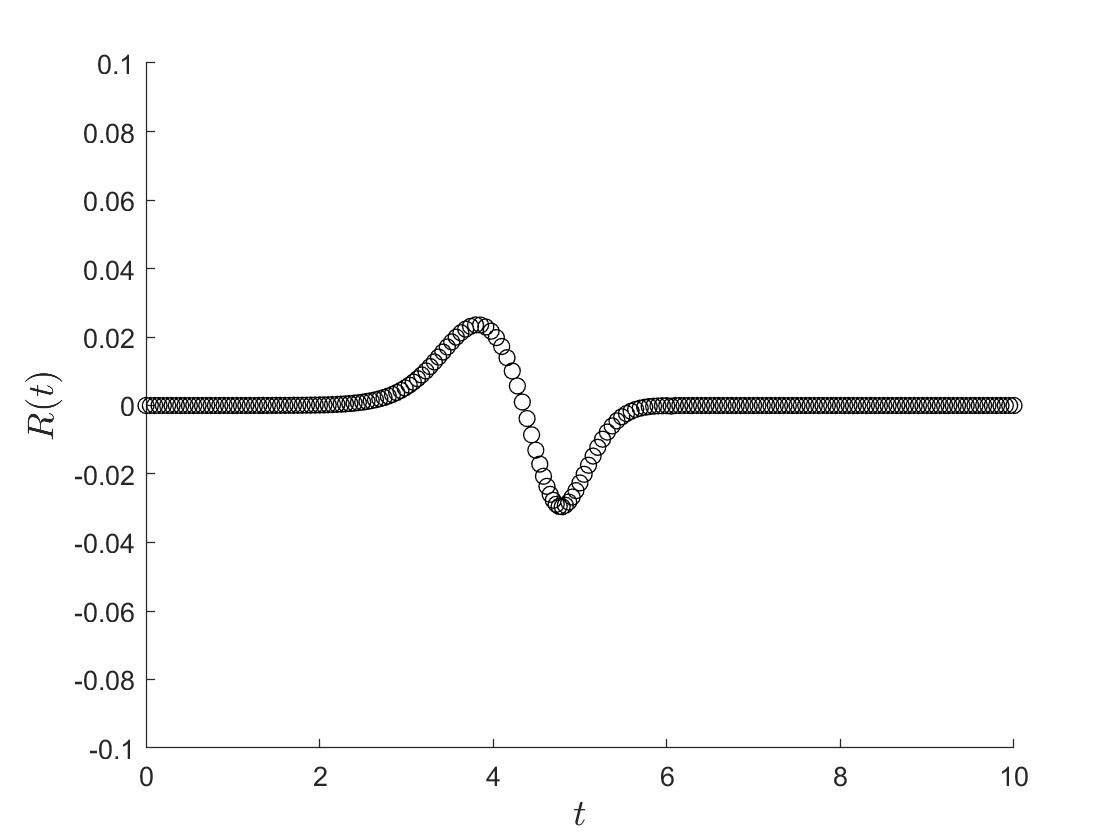}
            \caption{$m=2$}
            \label{m2_gvert}
        \end{subfigure}
       
        \begin{subfigure}{0.45\textwidth}
        \centering
            \includegraphics[width=\linewidth]{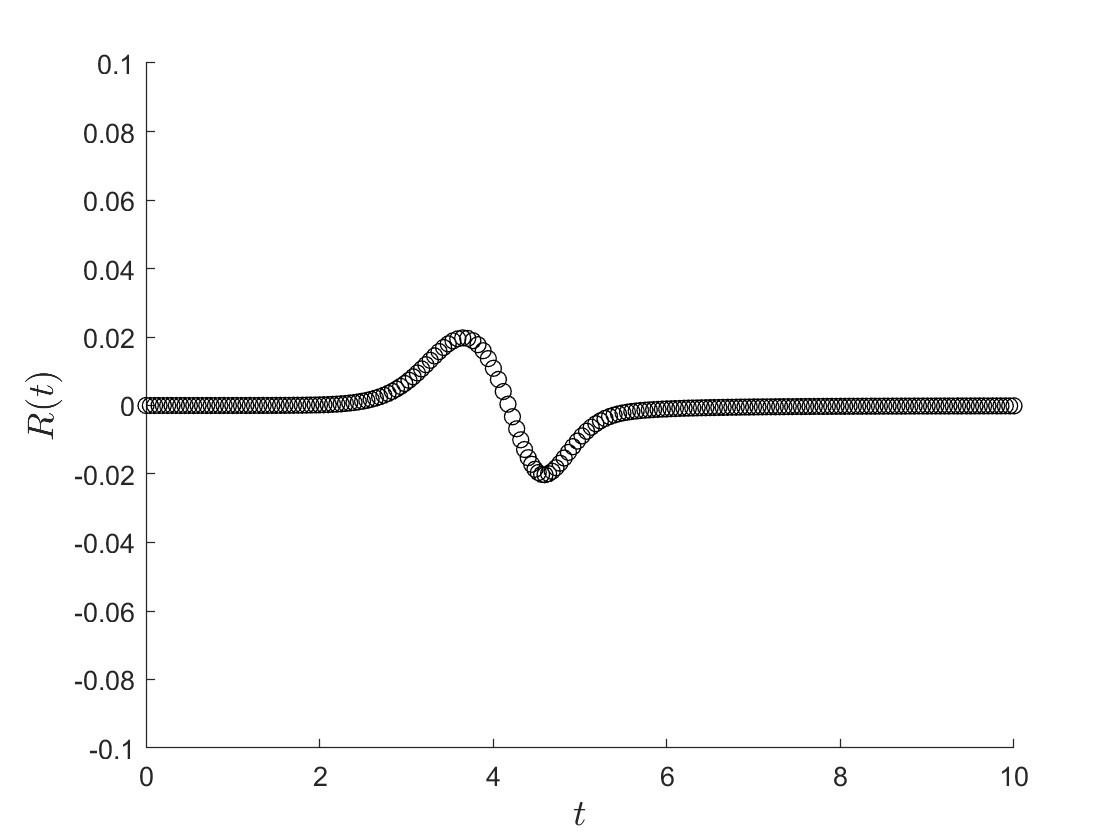}
            \caption{$m=3$}
            \label{m3_gvert}
        \end{subfigure}%
        \begin{subfigure}{0.45\textwidth}
        \centering
            \includegraphics[width=\linewidth]{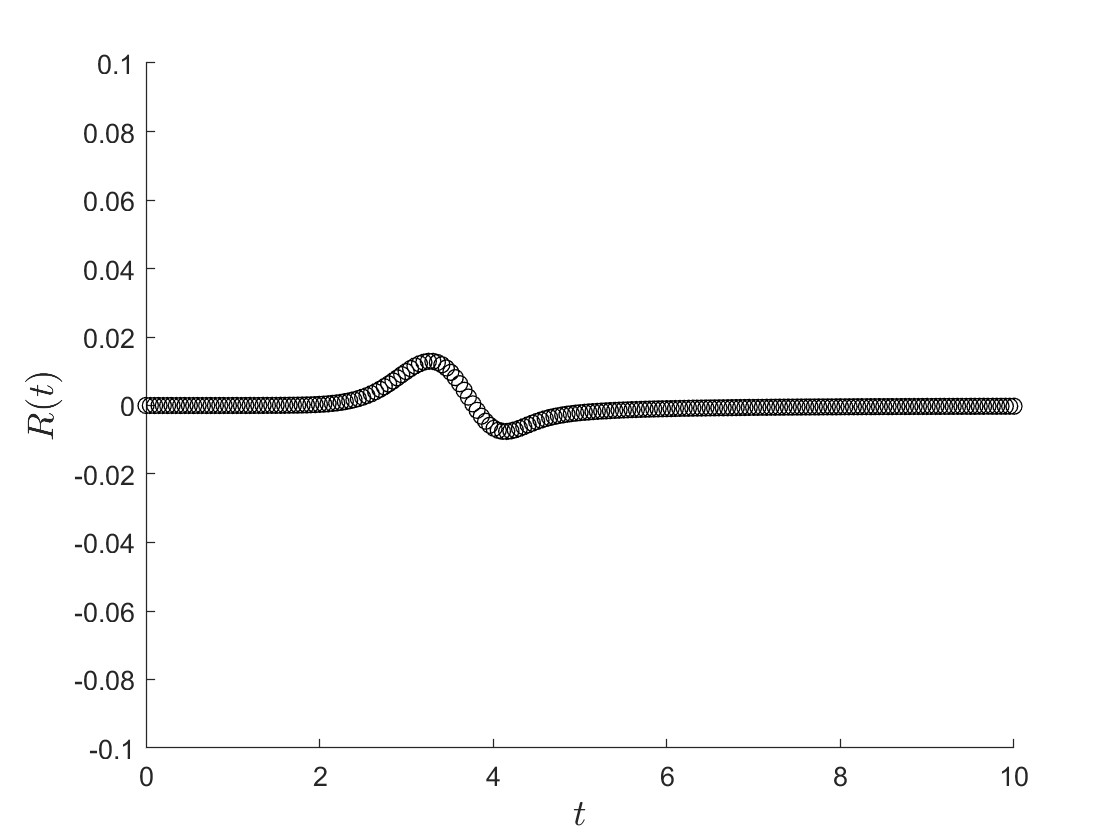}
            \caption{$m=\infty$}
            \label{minf_gvert}
        \end{subfigure}%
        \caption{Estimated vertical shift for an initial Gaussian wave displacement corresponding to various bathymetries, where $R(t) = -x_0(t)/\alpha$, where $\alpha = 1$.}
        \label{shorelineforgaussian}
        \end{figure}

In this section we numerically verify our method for recovering $\eta_0$ in cases where $m\in\left\{ 1, 2, 3,+\infty\right\}$, that is
for inclined parabolic  bays of different shapes and an infinite sloping beach.
In all of bathymetries we consider an ``$N$-wave''

\begin{equation}
\eta_0(x)=2.5\times 10^{-3}e^{-3.5(x-1.9625)^2}-1.25\times 10^{-3}e^{-3.5(x-1.4)^2},
\end{equation}
and a Gaussian wave
\begin{equation}\label{gaussian}
\eta_0(x)=5\times 10^{-3}e^{-(x-3)^2},
\end{equation}

with zero initial velocity. Existing code provided by Rybkin et al. \cite{Rybkin21} was used to generate the shoreline data. We then implemented \eqref{m=inf} and \eqref{m=2} respectively to recover the initial displacements. Comparison of the exact initial wave profiles and those predicted by our model can be seen in Figures \ref{numerical_gaussian} and \ref{numerical_nwave}.
Corresponding shoreline movements can be seen in Figures \ref{shorelineforgaussian} and \ref{shorelinefornwave}. It is worth noting that when we consider the same initial displacement for various bathymetries, the amplitude of the shoreline movement decreases as $m$ increases.

\begin{figure}[p]\begin{center}
    \includegraphics[height = 8cm]{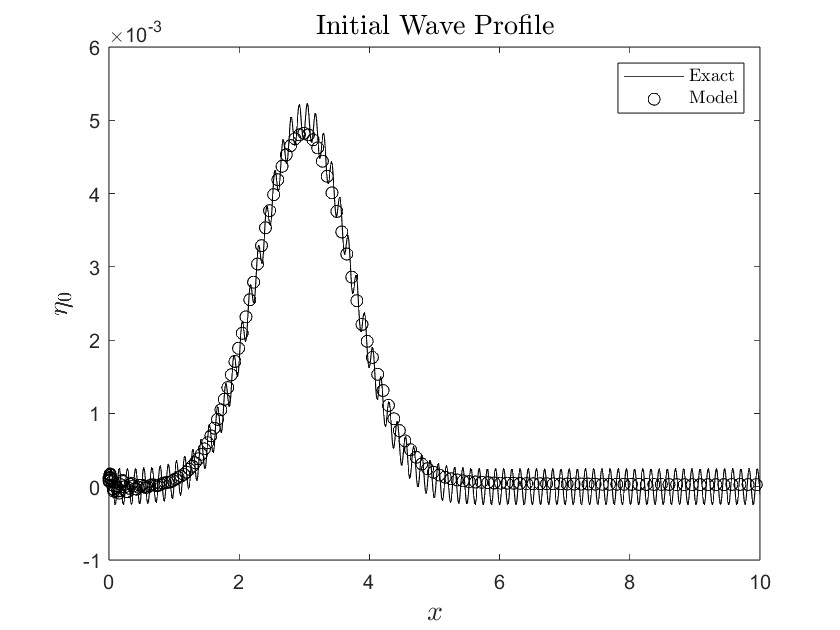} 
    \caption{Our model effectively cuts off any high frequency waves, as can be seen with a Gaussian initial wave profile ($m=\infty$).}\label{minf_noisegaussian}.
    \end{center}\end{figure}

It is common when a long tsunami wave is masked by wind waves that have higher frequency.
Typically,
the tsunami wave length is above 1 kilometre, while wind waves have length of 90 to 180 metres.
The integral transform we derived cuts off high-frequency oscillations.
To demonstrate that we consider a long wave with added disturbance

\begin{equation}
\eta_0(x) = 5\times 10^{-3} e^{-(x-2)^2}+2.5\times 10^{-4}\sin(50x).
\end{equation}
Using the same methodology we are again able to recover the initial wave profile (see Fig. \ref{minf_noisegaussian})  without the noise created by the high-frequency oscillations. While this is a coincidence of our numerical scheme, it holds practical significance. It suggests that our method can effectively reconstruct the long wave, presumably generated by the tsunami source, even in the presence of noise from common wind waves.

\begin{figure}[p]\begin{center}
  \includegraphics[height = 8cm]{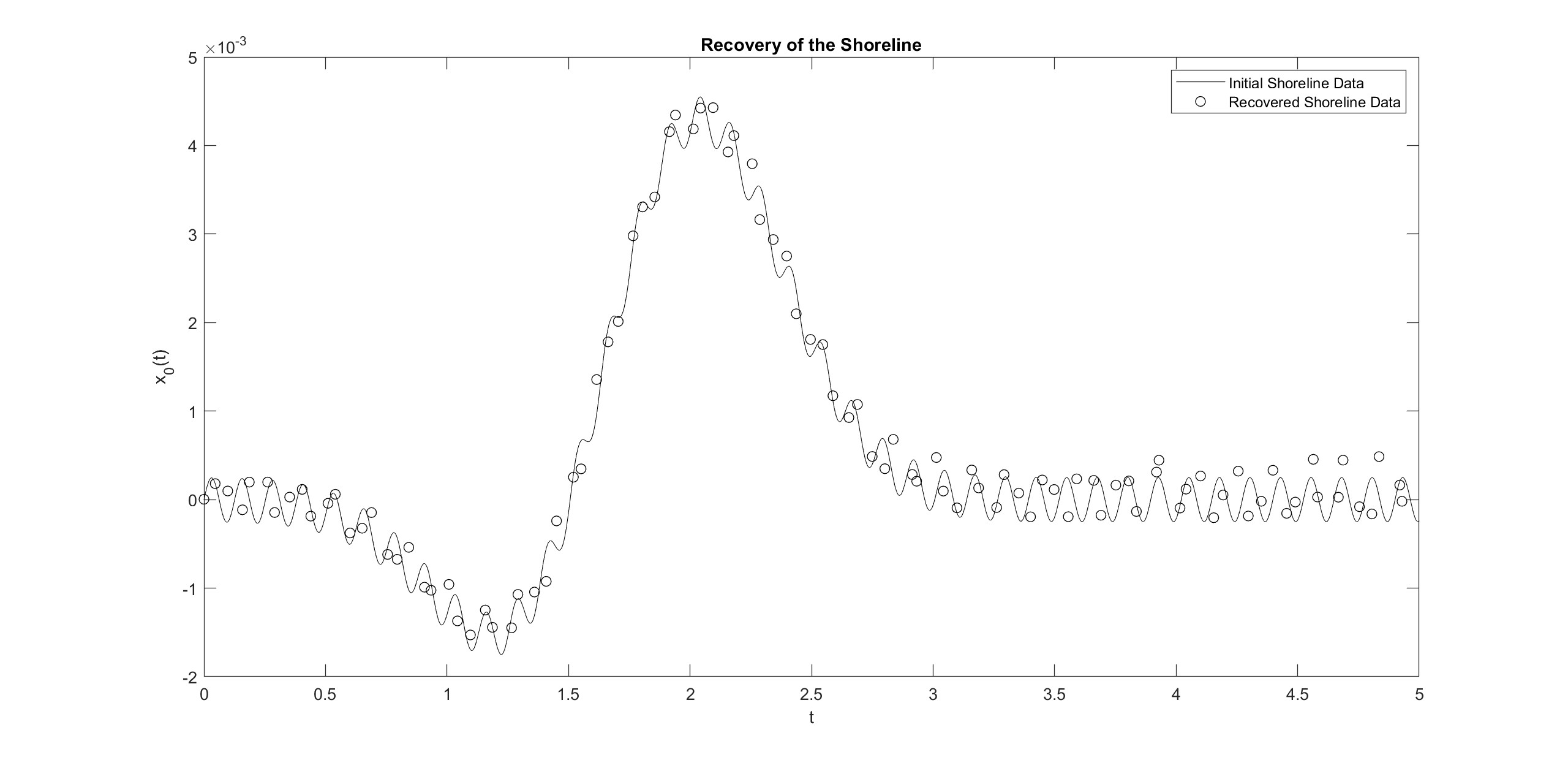}
  \caption{While our numerical model doesn't remove noise found within the shoreline data, the model is able to cope with the existence of noise and recover it's behaviour. In this case, our model begins by taking an analytical shoreline function with noise, given by $x_0(t) = 5\times10^{-3}e^{-3.5(t-1.9625)^2} - 2.5\times10^{-3}e^{-3.5(t-1.4)^2} + 2.5\times10^{-4}\sin(50t)$. This $x_0(t)$ is utilised to compute $\eta_0$ through the inverse problem. Subsequently, the model verifies the validity of this ensuring that the $\eta_0$ gained by recovering $x_0(t)$ through the direct problem directly matches the analytical $x_0(t)$ used to begin the computations. (Done in the case $m = 2$)}\label{gaussian_noise}
\end{center}\end{figure}

\begin{figure}[p]
\centering
\includegraphics[height=8cm]{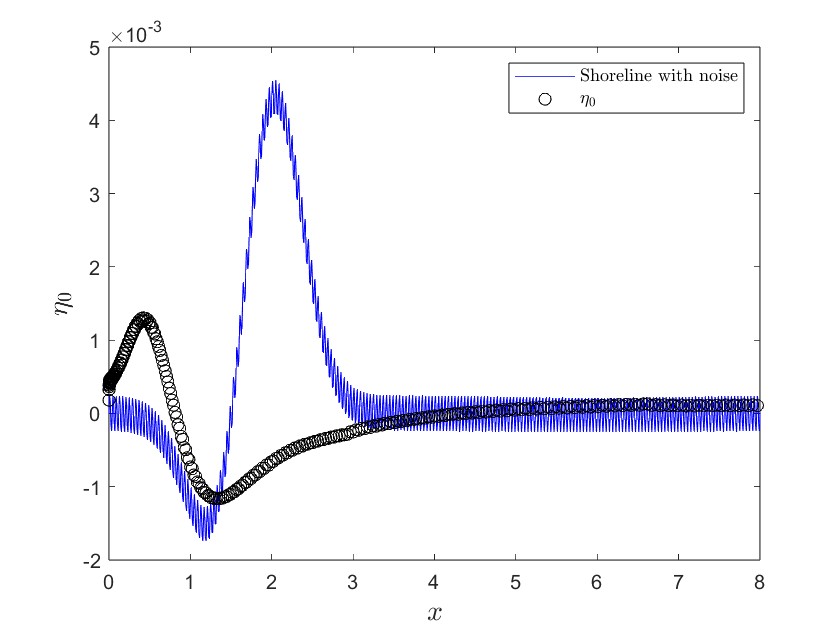}
\caption{The model appears to reduce high frequency noise. Here we took the shoreline to be $x_0(t)= 5\times10^{-3}e^{-3.5(t-1.9625)^2} - 2.5\times10^{-3}e^{-3.5(t-1.4)^2} + 2.5\times10^{-4}\pi\sin(50\pi t)$ and then computed $\eta_0$ using \eqref{m=inf} which recovered a relatively smooth curve, as seen above.}\label{shorelinenoise}
\end{figure}

For analogous reasons our model can handle noise in the shoreline data effectively (see Figures \ref{gaussian_noise} and \ref{shorelinenoise}). This capability may offer practical benefits, enhancing its robustness against noisy data collection at the shore.

It is worth noting that the noise reduction happens because of the numerical implementation of the integral. If one was able to compute the derived integral transform exactly, one would recover exact initial displacement with the noise.

\section{Conclusions}\label{sec:disc}
We have put forth and solved an inverse problem for non-breaking tsunami waves in power shaped bays assuming zero initial velocity. We have shown that for non-braking waves in power-shaped bathymetries it is possible to recover the initial displacement of the water from the shoreline oscillations under the assumption of zero initial velocity. 

Of course, the formulation of the problem is quite schematic from a practical standpoint: a one-dimensional problem, a bay with a constant slope, and a tsunami source relatively close to the coast. Nevertheless, this problem is part of the benchmarks for numerical tsunami models, and it is necessary for calculating possible scenarios to make predictive assessments of tsunami heights. To create tsunami hazard maps, calculations of potential tsunamis are currently being conducted using data from a synthetic earthquake catalogue, which includes approximately 100,000 events. Calculating such a large number of tsunamis from the source to the shore is too costly, so it is done in two stages: in the first stage, the characteristics of the tsunami are calculated up to depths of around 20-50 meters, and then the wave height is recalculated at the shore using various versions of Green's law \cite{sorensen2012probabilistic,baptista2017synthetic,basili2021making}.
This is where the non-linear problem of wave run-up on the shore can serve as an effective means to improve estimates obtained through Green's law, and it was used, for example, in tsunami hazard assessments in the Sea of Japan \cite{choi2011rapid}. The formula for the run-up height of a solitary wave \cite{Synolakis87} is especially often used. Refining the shape of the initial wave (and more generally, the initial flow velocity) in the run-up problem and assessing how critical it is for calculating run-up heights can be aided by solving the inverse problem.

While not considered here, we believe that a similar inverse problem can be treated in the case where the initial velocity is given as a function of the initial displacement, e.g., in the important case where $u_0=-(2\sqrt{x+\eta_0}-2\sqrt{x})$.
Indeed, our preliminary results show this to be possible in the plane beach bathymetry. We hope to return to this case in a future work.

Our results here consider a tsunami wave with source an arbitrary distance from the shoreline. However, this is a highly idealised situation. In practice dispersion can only be ignored when the wave is close to the shoreline. This suggests a more practical inverse problem where we have a finite bathymetry, that is $x\leq L$ for $L>0$ and attempt to recover the wave at $L$. This is a boundary value problem and the techniques developed in \cite{Antuono07} and \cite{Rybkin21} may be used to derive a shoreline equation.
Finally we note that our inversion method can be readily adjusted to the data read from a mareograph which is close to the shore. Indeed, using the angle of inclination the mareograph readings can be converted into a displacement of the shoreline.

\section{Acknowledgements}
This work was done as part of the 2023 summer REU program run by Dr.~Alexei~Rybkin and was supported by NSF grant DMS-2009980.
Dr.~Efim~Pelinovsky thanks a support from RSF 22-17-00153.
Oleksandr Bobrovnikov acknowledges support from Alaska EPSCoR NSF award \#OIA-1757348 and DMS-2009980.
We also thank  Dr.~Ed~Bueler for his valuable discussions of the problem with us. We also thank the UAF DMS for hosting us.
\newpage

\bibliography{final.bib}
\clearpage
\end{document}